\documentclass{article}

\usepackage{arxiv}

\usepackage[utf8]{inputenc} 
\usepackage[T1]{fontenc}    
\usepackage{hyperref}       
\usepackage{url}            
\usepackage{booktabs}       
\usepackage{amsfonts}       
\usepackage{nicefrac}       
\usepackage{microtype}      
\usepackage{lipsum}
\usepackage{graphicx}
\usepackage{multirow}
\usepackage{algorithm}
\usepackage{algorithmic}
\usepackage{amsmath}
\graphicspath{ {./images/} }

\title{Modification and Generated-Text Detection: Achieving Dual Detection Capabilities for the Outputs of LLM by Watermark}

\author{
 Yuhang Cai$^1$ \\
  School of Computer Science\\
  HeFei University of Technology\\
  HeFei  \\
  \texttt{2023110524@mail.hfut.edu.cn} \\
   \And
 Yaofei Wang$^2$  \\
   School of Computer Science\\
    HeFei University of Technology\\
 HeFei \\
  \texttt{wyf@hfut.edu.cn } \\
  \And
Donghui Hu$^3$  \\
  School of Computer Science\\
  HeFei University of Technology \\
  HeFei \\
  \texttt{hudh@hfut.edu.cn } 
\And
Chen Gu $^4$ \\
 School of Computer Science\\
  HeFei University of Technology \\
  HeFei\\
  \texttt{guchen@hfut.edu.cn } 
}

\begin{document}
\maketitle
\begin{abstract}
The development of large language models (LLMs) has raised concerns about potential misuse. One practical solution is to embed a watermark in the text, allowing ownership verification through watermark extraction. Existing methods primarily focus on defending against modification attacks, often neglecting other spoofing attacks. For example, attackers can alter the watermarked text to produce harmful content without compromising the presence of the watermark, which could lead to false attribution of this malicious content to the LLM. This situation poses a serious threat to the LLMs service providers and highlights the significance of achieving modification detection and generated-text detection simultaneously. Therefore, we propose a technique to detect modifications in text for unbiased watermark which is sensitive to modification. We introduce a new metric called ``discarded tokens", which measures the number of tokens not included in watermark detection. When a modification occurs, this metric changes and can serve as evidence of the modification. Additionally, we improve the watermark detection process and introduce a novel method for unbiased watermark. Our experiments demonstrate that we can achieve effective dual detection capabilities: modification detection and generated-text detection by watermark.
\end{abstract}

\keywords{LLM, LLM Watermark, Modification Detection, Robustness}

\section{Introduction}
The powerful generative capabilities of LLMs have greatly enhanced the human capabilities to create text. Whether in literary creation, news writing, or technical documentation, people can complete work more efficiently and quickly with the help of LLMs. 
However, this technological advance has also raised concerns about the abuse of LLMs. LLMs can generate creations that are difficult to distinguish from human works and potentially create misleading statements or false information\cite{pan_risk_2023,kim2024llmsonlineemergingthreat}. Therefore, implementing practical detection tools to determine whether a text is generated by AI becomes particularly important \cite{pmlr-v235-chakraborty24a,Mitchelldetectgpt}.

Watermark is a promising method to reduce the risks of LLM abuse \cite{kirchenbauer2023reliability,li2024statistical,Wu2023ASO,kamaruddin_review_2018,yoo_advancing_2024,yang2022tracing}. Previous watermarking methods\cite{kirchenbauer23a,zhao2024provable,lee2024wrote,wu2024dip,kuditipudi2024robust,pmlr-v247-christ24a,hu2024unbiased}  often identify machine-generated text based on statistics, which counts the number of tokens with watermark and compares it with the threshold to obtain the detection result. These methods can achieve strong robustness, as it is difficult for attackers to reverse the results of statistical detection by modifying a few tokens.

\begin{figure}[h]
    \centering
    \includegraphics[width=0.7\linewidth]{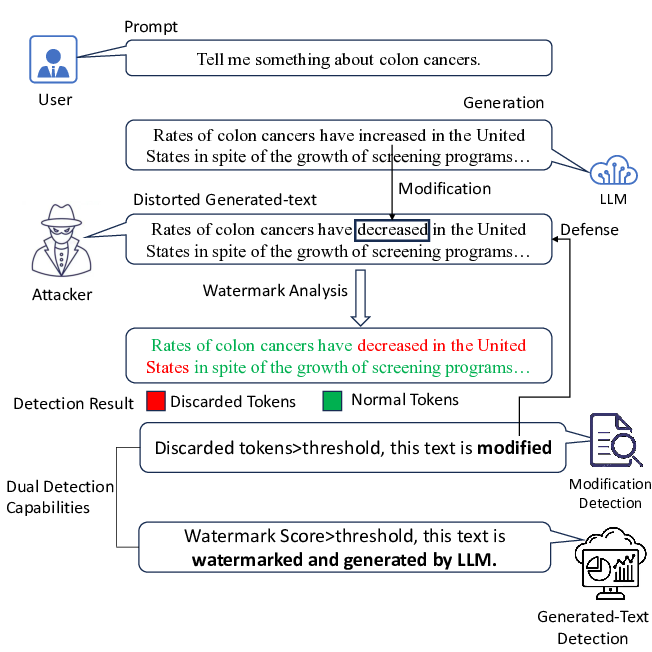}
 \caption {  \centering{The framework of dual detection capabilities for LLM-generated text by watermark. We analyze the discarded token(s) caused by modification, which fails to function as evidence for watermark detection. If the number of these tokens is larger than the threshold, it confirms the existence of modification. Meanwhile, we achieve generated-text detection by remaining tokens with watermark in the text.}}
    \label{fig:robust-fragile}
\end{figure}

The former watermarking methods do solve the problem of detecting machine-generated text. However, their detection strategy is insensitive to tiny modifications, and this advantage triggers another potential abuse. For instance, the text is vulnerable to tampering, leading to ambiguity or a direct distortion of the intended meaning. As shown in TABLE \ref{tb-examples}, watermarked texts can still have the watermark even after modification and are detected as machine-generated according to ``unpublished'' \cite{pang2024freelunchllmwatermarking}. Modified watermarked text may be spread through the internet and then used as evidence, which could falsely attribute harmful content to the LLMs mentioned by Gloaguen in ``unpublished'' \cite{gloaguen2024discoveringcluesspoofedlm}, making the company a scapegoat. This kind of spoofing attack utilizes the strength of the robust watermark while the former watermark is far from enough to defend it because these methods often ignore the necessity of modification detection for generated texts and fail to achieve dual detection capabilities by watermark: modification detection and generated-text detection.

To achieve dual detection capabilities by watermark as shown in Fig. \ref{fig:robust-fragile}, we propose the concept of modification detection based on watermark for LLMs. By analyzing existing watermarking methods, we find unbiased watermark $\delta$-reweight method \cite{hu2024unbiased} is sensitive to modification and has the potential to achieve modification detection. If a token is modified, the next several tokens with $\delta$-reweight watermark are inconsistent with the tokens sampled by LLM and become discarded tokens that fail to function as evidence for watermark detection. We call this situation inconsistent distortion and design a novel modification detection method named inconsistent diffusion detection (IDD) to determine whether the watermarked text has been modified. As shown in Fig. \ref{fig:robust-fragile}, any tiny modification can influence several tokens with watermark, causing inconsistent distortion and obvious changes on discarded tokens. Then compare the number of discarded tokens to the threshold to confirm the occurrence of modification. Experiment shows that IDD can achieve high accuracy in detecting modifications such as addition, deletion, and replacement. Besides, the unaffected watermark can still be extracted to confirm the generated text. As illustrated in TABLE \ref{tb-examples}, with the help of dual detection capabilities, our detection method can simultaneously detect whether the text is generated and whether the generated text has been modified. Meanwhile, we improve the original maximin variant of the log-likelihood ratio score (mmLLR) \cite{hu2024unbiased} and introduced a new watermark detection score named drLLR, which is calculated by dropping the abnormal score of tokens. Experiments show that the drLLR method shows satisfactory performance in watermark detection accuracy and robustness.
\begin{table*}[h]
\centering
\caption{\centering{Examples of spoofing attacks on LLM-generated watermarked text and corresponding results}}
\belowrulesep=0pt
\aboverulesep=0pt
\begin{tabular}{ll|p{7cm}|p{2cm}|p{2cm}}
 \toprule
\multicolumn{2}{l|}{Type}                                  & Watermarked text                                                                                   & Previous detection methods             &Our dual detection method                                                                                     \\ \midrule
\multicolumn{2}{l|}{Original}                              & A clinical trial showed a relation between vitamin D insufficiency and increased morbidity...      & \multirow{4}{*}{Watermarked} & Watermarked                                                                               \\ \cline{1-3} \cline{5-5} 
\multicolumn{1}{l|}{\multirow{3}{*}{\begin{tabular}[c]{@{}l@{}}Modification\\ attack\end{tabular}}} & Deletion    & A clinical trial showed a relation between vitamin D {\textbf{insufficiency}} and increased morbidity...                    &                              & \multirow{3}{*}{\begin{tabular}[c]{@{}l@{}}Watermarked \\ and \textbf{modified}\\\end{tabular}} \\ \cline{2-3}
\multicolumn{1}{l|}{}                        & Replacement & A clinical trial showed \textbf{{a} \underline{no}} relation between vitamin D insufficiency and increased morbidity...     &                              &                                                                                           \\ \cline{2-3}
\multicolumn{1}{l|}{}                        & Addition    & A clinical trial showed a \textbf{\underline{weak}} relation between vitamin D insufficiency and increased morbidity... &                              &                                                                                           \\ \bottomrule
\end{tabular}
\label{tab:table1}
\label{tb-examples}
\end{table*}

Our main contributions are summarized as follows:
\begin{enumerate}
    \item We first propose the concept of modification detection to defend against potential spoofing attacks, which are often ignored by the former methods and short of the corresponding metrics. 
    \item We design a modification detection method named IDD for LLMs based on the characteristic of $\delta$-reweight. By calculating the number of discarded tokens, the method achieves accurate modification detection. 
    \item We propose a novel watermark detection score drLLR for $\delta$-reweight. With IDD and drLLR method, we achieve dual detection capabilities: modification and generated-text detection for output of LLM by watermark.
\end{enumerate}

\section{Related Works}

\subsection{Watermarking for LLM}

Kirchenbauer et al.\cite{kirchenbauer23a} introduced a pioneering watermarking framework tailored for LLMs that embeds watermarks with minimal text quality impact. It creates ``green" token list randomly before generating tokens, and encouraging the model to choose from them. This type of watermarking method increases robustness, but it disrupts the output distribution of the model\cite{singh2023newevaluationmetricscapture}. To reduce the impact on the quality of text generation, Lee\cite{lee2024wrote} considered text entropy to modify logits adaptively. To enhance the robustness of the watermark, Zhao\cite{zhao2024provable} improved the robustness by fixing the division of the red and green lists. However, the above methods cannot avoid affecting the quality of the text generated by the LLM. To ensure the quality of text generation, Hu et al.\cite{hu2024unbiased} and Wu et al.\cite{wu2024dip} proposed unbiased watermark algorithms that can maintain probability distribution while embedding watermark.

\section{Method}

\subsection{Preliminary}

In LLMs generation, $P_M$ represents the probability distribution generated by pre-trained LLM, and $\mathcal{V}$ is the overall vocabulary set. In a typical LLM generation task, LLM receives a prompt $x_{-n_p:0}$ and outputs a sequence $x_{1:n}$ according to the prompt and the generated tokens $x_{-n_p: i-1}$ by gradually generating the next token $x_i$. When generating the token $x_i$, the probability of the token in the vocabulary set $\mathcal{V}$ is given by the conditional probability distribution $P_M (x_i\mid x_{-n_p: i-1})$. 
 
When embedding watermark by $\delta$-reweight\cite{hu2024unbiased}, the output probability distribution is adjusted from $P_M (x_i\mid x_{-n_p: i-1})$ to $P_{M,w} (x_i\mid x_{-n_p: i-1},\theta_i)$. The cipher $\theta_i$ is usually generated by a secret key $k \in \mathcal{K}$ and a fragment of the previous context, named \textit{texture key}, $ct$. Each $\theta_i$ is independent and follows the same distribution $P_{\Theta}$.

\subsection{Observations of Discarded Tokens}

There is no existence research for modification detection on text by watermark. To achieve modification detection, we must observe the significant differences in the generated watermarked text before and after modification. One type of difference can be reflected in discarded tokens, which are not involved in detection and fail to function as evidence for detection results. However, for watermarks that pursue robustness \cite{kirchenbauer23a,lee2024wrote,zhao2024provable}, they are insensitive to modification because tokens with robust watermark are hard to be affected by modified tokens, resulting in inapparent difference because the number of discarded tokens increases slowly. Therefore, they are not suitable for modification detection. Meanwhile, we turn to another type of watermark: watermarks that are sensitive to modification. When face modification, the modified token influences the next several tokens even if these tokens are not modified, which causes obvious changes on discarded tokens and function as evidence for the existence of modification. 

\subsection{Inconsistent Distortion}

\begin{figure}[h]
    \centering
    \includegraphics[width=0.7\linewidth]{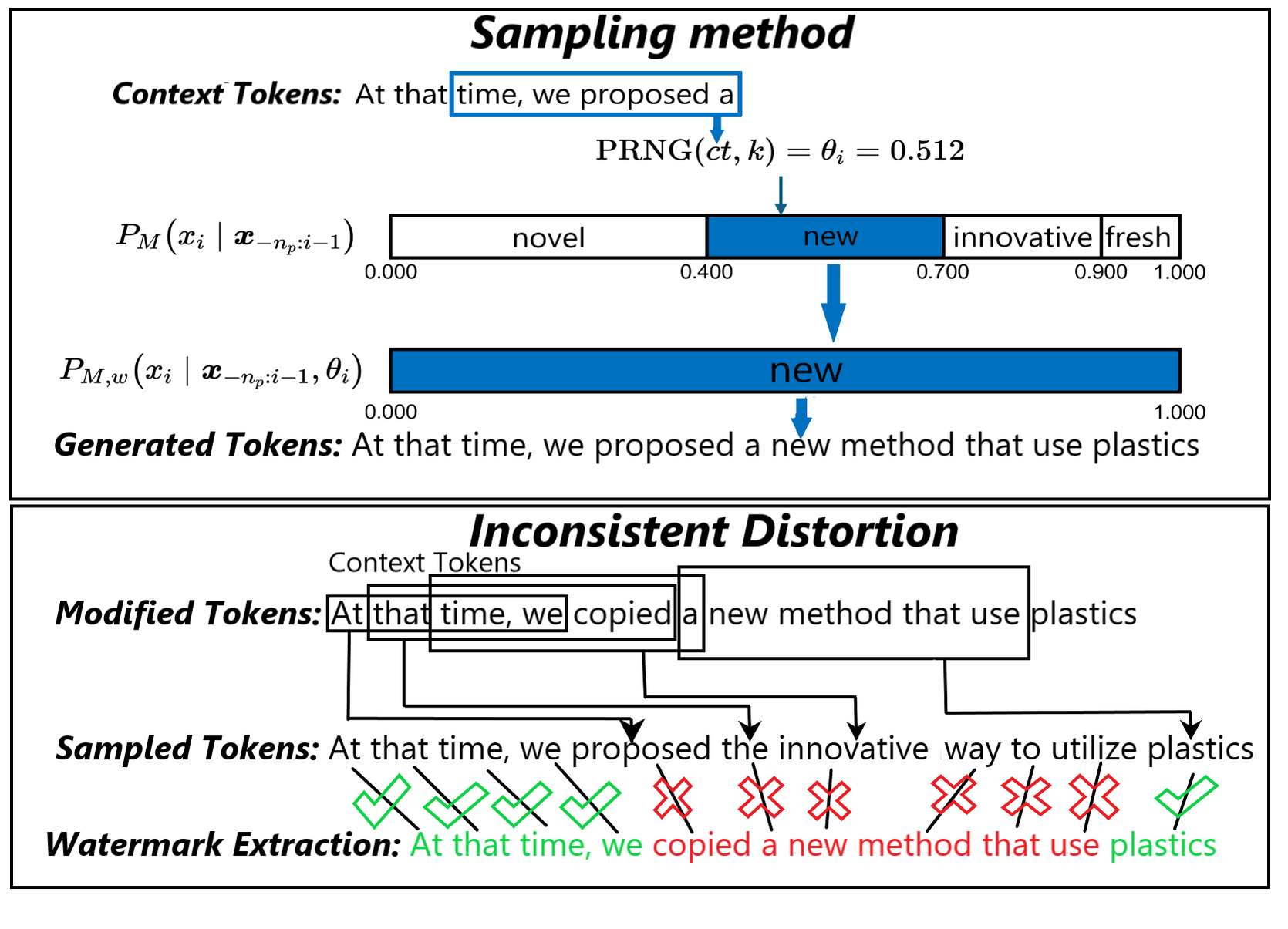}
    \caption{\centering{Sampling method of $\delta$-reweight and inconsistent distortion of $\delta$-reweight caused by modified token(s). The upper part of the figure illustrates the sampling process without a modified token. The lower part demonstrates the process that modified token in context tokens disturbs sampling method and result in inconsistent tokens marked in red until there is no modified tokens in context tokens. The unaffected tokens in green are still consistent with sampled tokens and function as evidence for detection result.}}
    \label{fig:fragility}
\end{figure}

The $\delta$-reweight \cite{hu2024unbiased} method is an unbiased watermarking method. As shown in the upper part of Fig. \ref{fig:fragility}, by using the key $k$ and the specified context parameter $n_{ct}$ (default to be the most recent five tokens) in the hash function, the seed of the pseudo-random number generator (PRNG) is obtained after hashing, and a pseudo-random number $\theta_i$ is generated.  Based on the probability distribution, if a token is selected, the output probability of the token is set to 1 while probability of other tokens is set to 0. This sample method is unbiased, also sensitive to modification for its legal token selection.

Specifically, given a LLM $M$ and the corresponding generated context $x_{-n_p:i-1}$. The probability of normal generation and the probability of containing watermark of the token $x_i$=``new" are $P_M (x_i = \text{``new"} )$ and $P_{M,w}\left(x_i = \text{``new"} \right)$, respectively. For example $P_{M,w}\left(x_i = \text{``new"} \right)=1$ according to  $\delta$-rewight while $P_M (x_i = \text{``new"} ) = 0.3$ as shown in the top of  Fig. \ref{fig:fragility}. Therefore the legal token $x_i$ generated based on watermarked LLM can only be ``new". If the context tokens before ``new" are modified while other tokens remain unchanged, $x_i'$  in text is ``new" but $P_{M,w}\left(x_i = \text{``new"} \right)=0$. The token ``new" in the text is inconsistent with the token ``innovative" that should be generated at that position according to context tokens and fails to function as evidence for watermark detection. As shown in the bottom of Fig. \ref{fig:fragility}, the token ``new" becomes an inconsistent token for $\delta$-reweight which is also a discarded token not included in watermark detection. This situation occurs until there are no modified tokens in context tokens and makes obvious changes in inconsistent tokens that are able to reflect modification accurately. So $\delta$-reweight has the potential to achieve modification detection and we call this kind of characteristic inconsistent diffusion caused by $\delta$-reweight sampling method. Then we build the modification detection method named inconsistent diffusion detection (IDD) based on this characteristic.

\subsection{Modification Detection}

We described the modification detection method IDD in Algorithm \ref{algorithm1}. Modification score $n_{it}$ is introduced to represent the number of inconsistent tokens that $x_i'\ne x_i$, which means token $x_i'$ in the text is inconsistent with token $ x_i$ sampled by watermark method. According to $\delta$ method, we constructed the same sampling function as the generation to restore the scene at the time of generation. Only one watermarked token can be generated based on sampling functions, and tokens different from the watermarked token can be classified as inconsistent tokens. We judge the tokens by $f_{it}(x_i)$ as follows:
\begin{equation}
    f_{it}(x_i)=\left\{\begin{matrix} 
 1,   \ if \ x_i' \ne x_i\\  
  0,    \ if \ x_i'= x_i
\end{matrix}\right. 
\end{equation}
where $1$ represents an inconsistent token, and 0 confirms a consistent token. 
We accumulate the number of  inconsistent tokens to get the final result $n_{it}$:
\begin{equation}
    n_{it}=\sum_{i=1}^{N}f_{it}(x_i)
\end{equation}

Then, we set a threshold. If $n_{it} $ is larger than the threshold, confirm the modification. 

\begin{algorithm}[tb]
\caption{Modification Detection and Watermark Detection}
\label{algorithm1}
\textbf{Input}: LLM $M$, generated text $x$, text length $n$, modification threshold $M_T$

\textbf{Output}: Average watermark detection score $drLLR_{avg}$, Inconsistent token number $n_{it}$, Modification flag $M_{flag}$


\begin{algorithmic}[1] 
\STATE Let drLLR$=0$, $M_{flag}$=false, $i$=1, $n_{it}=0$, $t = 0$
\WHILE{$i$ $in $ $n$}
\STATE Sample the token $T_i$ according to $\delta$-reweight
\IF {$T_i$ equals $x_i$ }
\STATE $drLLR(x_i)$=$\log$(1/$P_{M}(x_i \mid \boldsymbol{x}_{-n_p: i-1}) $)
\STATE Add $drLLR(x_i)$ to $drLLR$
\ELSE
\STATE $n_{it}=n_{it}+1$
\ENDIF
\STATE $i=i+1$
\ENDWHILE
\STATE $drLLR_{avg} = \frac{drLLR}{n}$
\IF {$\frac{n_{it}}{n_{ct}}>M_T$ }
\STATE $M_{flag}=true$
\ENDIF
\STATE \textbf{return} $drLLR_{avg}$, $n_{it}$, $M_{flag}$
\end{algorithmic}
\end{algorithm}

\subsection{Enhance Watermark Detection}

For the watermarked text generated by $\delta$-reweight, the inconsistent tokens significantly affect the detection score of the original method LLR and fail detection. To enhance the robustness and mitigate the impact of inconsistent tokens on watermark detection, we use an improved LLR for watermark extraction, named drLLR, as shown in Algorithm \ref{algorithm1}. Traditional LLR score results in $LLR(x_i)=-\infty$ when the token $x_i$ is modified, then makes the detection fail. However, if a token is modified and results in a negative infinite score, it should not be used as a detection basis but discarded directly. Therefore, we set the score of inconsistent tokens to 0 to drop these tokens:
\begin{equation}
    drLLR(x_i)=\left\{\begin{matrix} 
  \log \frac{P_{M, w}\left(x_i \mid \boldsymbol{x}_{-n_p: i-1}, \theta_i\right)}{P_M\left(x_i \mid x_{-n_p: i-1}\right)},   \  if \ x^{'}_i=x_i\\  
  0,    \ if \ x^{'}_i\ne x_i
\end{matrix}\right. 
\end{equation}
where $x_i'\ne x_i$ indicates that token $x_i'$ in the text is inconsistent with token $ x_i$ sampled by watermark method.

We only calculate the sum of the score of other tokens and divide it by the total number of tokens to reflect the strength of the watermark. If the average score drLLR is greater than the threshold, we detect the watermark in the text and identify that LLMs generate the text. Otherwise, we consider that the text does not carry the watermark:
\begin{equation}
    drLLR_{avg}=\frac{1}{N} \sum_{N}^{i=1} drLLR(x_i)
\end{equation}

\section{EXPERIMENTS AND RESULTS }

\begin{figure*}[h]
	
	\begin{minipage}{0.51\linewidth}
		\vspace{3pt}
		\centerline{\includegraphics[width=\textwidth]{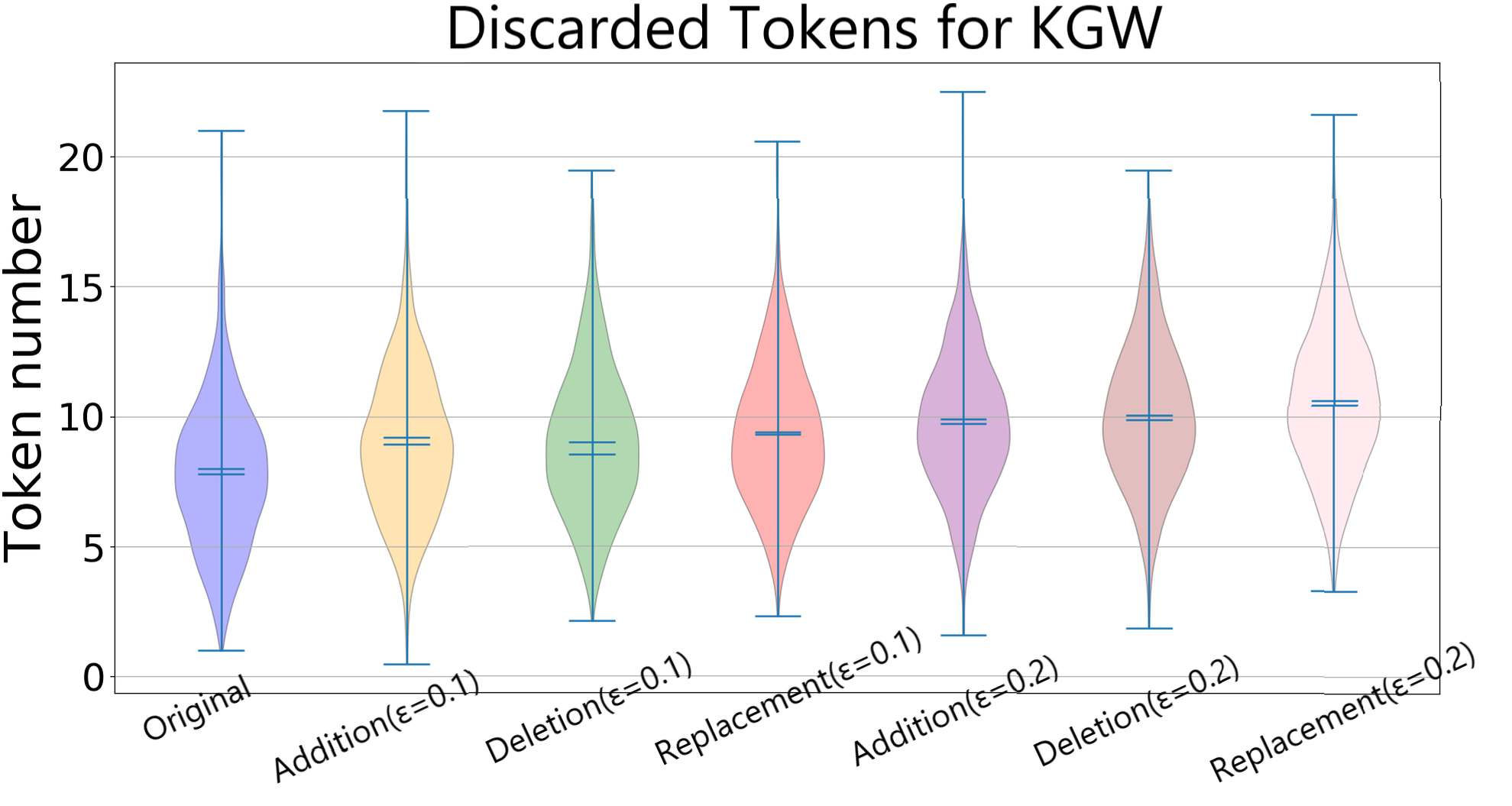}}
		\centerline{(a)}
	\end{minipage}
	\begin{minipage}{0.49\linewidth}
		\vspace{3pt}
		\centerline{\includegraphics[width=\textwidth]{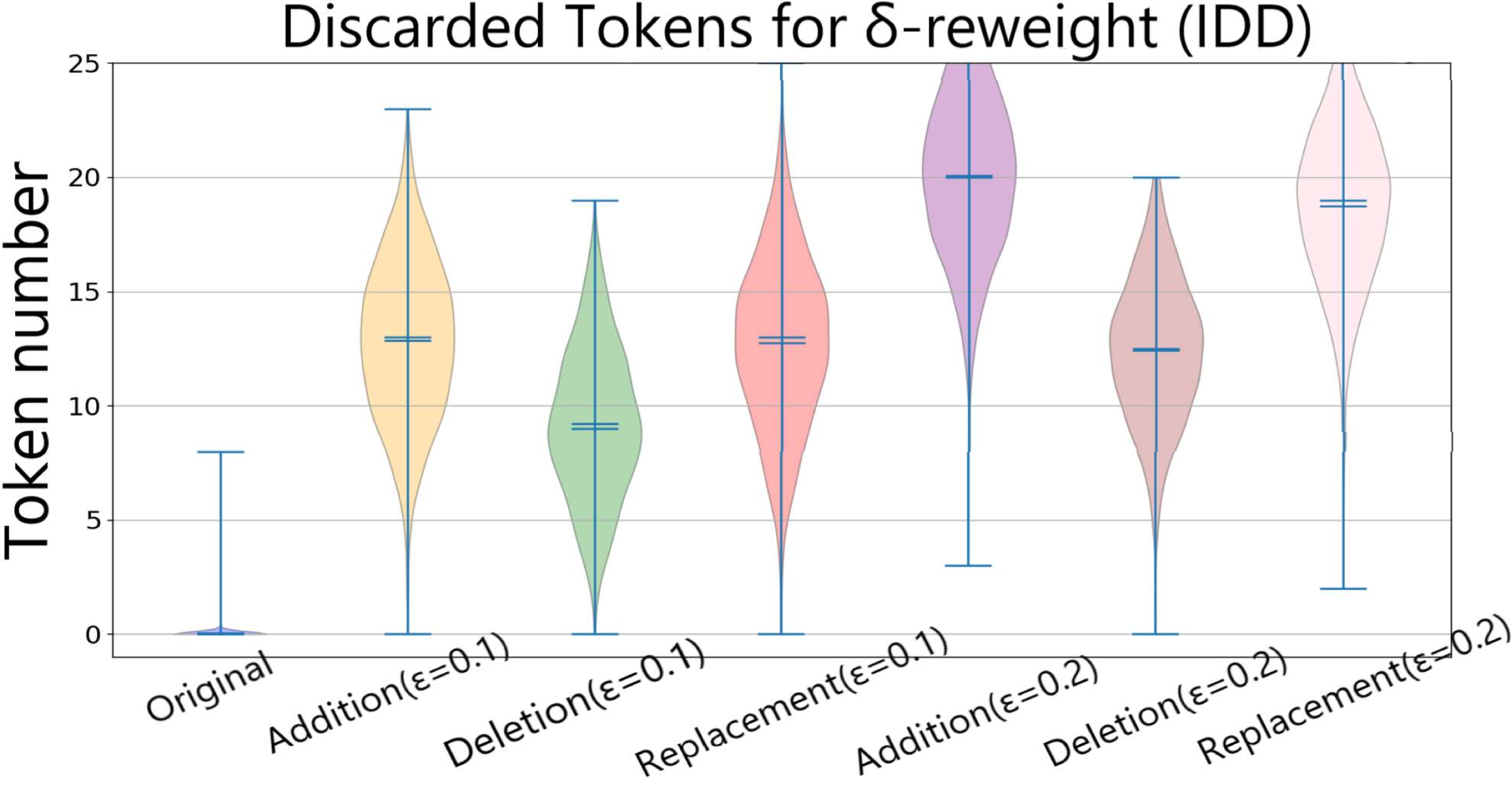}}
	 
		\centerline{(b)}
	\end{minipage}
 
	\caption{Distribution of the number of discarded tokens (tokens in red list for KGW and inconsistent tokens for $\delta$-reweight) in text under different attack. }
	\label{fig:kgw-idd}
\end{figure*}

To evaluate the effectiveness of our method, we conduct experiments for the outputs of LLM with watermark in dual detection capabilities: modification detection and generated-text detection. 
Both evaluations are under a low-entropy environment because modification detection is worthy, especially in low-entropy circumstances like Medical Q\&A, Scientific Q\&A, etc.

\subsection{Settings and Datasets }

To simulate a realistic low-entropy environment, we obtain 1,000 questions from the PubMedQA dataset \cite{jin2019pubmedqa} and use them as prompts. We use the model OPT-6.7B proposed by Zhang ``unpublished'' \cite{zhang2022opt} and set the sampling method to Top-$p$ ($p=0.9$) and Top-$k$ ($k$=50) to generate two datasets, one without watermark and the other with watermark, each containing 1000 pieces of text and max length of each text is 30. We choose three types of modification that are commonly used in actual use: addition, deletion, and replacement. We use a random perturbation parameter $\epsilon$ to create datasets with different modification strengths. For example, $\epsilon$ = 0.1 means 10\% of tokens are modified in the datasets. 

\subsection{Baseline and Evaluation Metrics}

\textbf{Baseline} For modification detection, we improve the KGW method\cite{kirchenbauer23a} and use it as the baseline. We implement discarded tokens (which means these tokens are not involved in detection and fail to function as evidence for detection result) in KGW by counting the number of tokens in the red token set and Z-score, which reflects the number of red tokens to detect modification, then evaluate the performance of modification detection.  Besides, we build the test based on a hypothesis: modifications damage part of watermarks, but there are enough identifiable watermarks to detect. If $z-score < \left | threshold \right |$, we report watermarked but modified. For generated-text detection, we choose KGW and $\delta$-reweight\cite{hu2024unbiased} with the original LLR method maximin variant of the LLR (mmLLR) as watermark baselines. Specifically, we set KGW with a fixed green list proportion $\gamma$ = 0.5 and diverse logit increments $\delta$ =1. The hyper-parameter grid\_ size of mmLLR is set to 10.

\begin{table*}
    \centering
    \caption{\centering{Results of modification detection under different perturbation strength and attacks.}}
    \belowrulesep=0pt
\aboverulesep=0pt
    \begin{tabular}{c|cccc|cccc} 
    \toprule
\multirow{2}{*}{Datasets}& \multicolumn{4}{c}{KGW (Improved)}& \multicolumn{4}{|c}{$\delta$-reweight (IDD)}\\
         \cline{2-9}& TPR & FPR  & Recall&F1-score& TPR & FPR  & Recall&F1-score\\\midrule 
         Addition($\epsilon$=0.1)& 0.370& 0.630& 0.601&0.458& 0.997& 0.003& 0.997&0.998\\  
         Addition($\epsilon$=0.2)& 0.508& 0.492& 0.674&0.580& 1.000& 0.000& 0.996&0.998\\ \midrule
         Replacement($\epsilon$=0.1)& 0.392& 0.608& 0.615&0.479& 0.990& 0.010& 0.990&0.995\\ 
         Replacement($\epsilon$=0.2)& 0.591& 0.409& 0.707&0.644& 0.996& 0.004& 0.996&0.998\\ \midrule
 Deletion($\epsilon$=0.1)& 0.449& 0.551& 0.647& 0.530& 0.997& 0.003& 0.997&0.998\\ 
 Deletion($\epsilon$=0.2)& 0.558& 0.442& 0.695& 0.619& 1.000& 0.000& 0.996&0.998\\ \bottomrule
    \end{tabular}
   \label{tab:modification result}
\end{table*}

\textbf{Evaluation Metrics} Currently, there are no studies about modification detection based on watermark for LLM-generated text, so we design our own watermark modification detection indicators. We compute TPR, FPR, Recall, F1-score to evaluate the ability of modification detection on different datasets. Especially, we set watermarked and modified text as positive examples, while watermarked and non-modified text as negative examples. For generated-text detection, we evaluated the performance of different watermarking methods on watermark strength. We report the Area Under Curve (AUC) of watermark detection. 

\subsection{Modification Detection Studies}

As shown in Fig. \ref{fig:kgw-idd}, after different types of attacks, the number of discarded tokens (in the red list) from the KGW method remains relatively unchanged, meaning this method is insensitive to modification. Conversely, discarded tokens (inconsistent tokens) from $\delta$-reweight with IDD method is nearly 0 on the unaltered dataset, but it rapidly increases on the tampered dataset, showing a significant difference. This illustrates the potential of achieving modification detection: $\delta$-reweight watermark is easily broken after being tampered with and affects the extraction of subsequent tokens, causing obvious changes in discarded tokens. For $\delta$-reweight, the impact can be reported by the number of inconsistent tokens, demonstrating the feasibility of using the IDD method to detect modification effectively.

	
	 
	 
 
TABLE \ref{tab:modification result} presents the detection results of the modification detection method on different datasets. IDD method exhibits outstanding performance in detecting modified text, while the improved KGW method cannot achieve accurate modification detection. Compared to the robust KGW method, $\delta$-reweight is more susceptible to modification and creates inconsistent distortion that is obvious and easy to detect. Therefore, by scanning sentences and identifying inconsistent tokens, the detector can confirm the occurrence of modification. As the intensity of modification increases, the accuracy of detection also improves. Additionally, IDD method is equally effective for tamper-free text, as it only needs to detect whether all tokens are intact to confirm that the sentence has not been tampered with. In short, IDD method utilizes the characteristic of inconsistent distortion and shows outstanding performance in modification detection.

\subsection{ Experiments for Generated-Text Detection }

Apart from modification detection, generated-text detection is an indispensable function for watermarks. So we show the result of AUC under different perturbation strength and perturbation methods in the Table \ref{tab:wm detect}. The drLLR method performs well in low entropy scenarios, achieving an AUC of 0.98 on the original watermarked dataset, surpassing both the KGW and mmLLR method (orignal LLR method). Compared to the KGW method, the AUC of drLLR slightly decreases when the attack strength reaches 0.1, and the decrease accelerates when it reaches 0.2. This is attributed to the characteristic of inconsistent distortion: the watermark carried by the text is prone to be broken, affecting the extraction of watermark on subsequent tokens, resulting lower detection score. Therefore, its robustness against modification is challenging to match the watermark that pursues robustness, yet it still achieves a certain level of robustness under modification attacks. This demonstrates that our detection method achieves dual detection capabilities for the output of LLM by effectively detecting modification in text and exhibiting certain robustness against modification.

\begin{table}
\centering
\caption{\centering{AUC of generated-text detection for different methods under different perturbation strength}}
\belowrulesep=0pt
\aboverulesep=0pt
\begin{tabular}{c|c|c|c|c}
\toprule
                                 Strength& Method & Addition & Replacement & Deletion \\ \midrule
 \multirow{3}{*}{$\epsilon$=0.0}& KGW& 0.960& 0.960& 0.960\\   & $\delta$-reweight (mmLLR)& 0.860& 0.860& 0.860\\ 
  & $\delta$-reweight (drLLR)& 0.989& 0.989&0.989\\ \midrule
 \multirow{3}{*}{$\epsilon$=0.1}& KGW& 0.925& 0.924& 0.933\\  & $\delta$-reweight (mmLLR)& 0.681& 0.657& 0.671\\ 
  & $\delta$-reweight (drLLR)& 0.954& 0.943&0.953\\ \midrule
 \multirow{3}{*}{$\epsilon$=0.2}& KGW& 0.915& 0.868& 0.892\\ & $\delta$-reweight (mmLLR)& 0.580& 0.562& 0.566\\ 
  & $\delta$-reweight (drLLR)& 0.890& 0.751&0.797\\ 
  \bottomrule
\end{tabular}
\label{tab:wm detect}
\end{table}

\section{Conclusion}

We propose a novel conception of modification detection based on watermark for LLM-generated text to defend against spoofing attacks. We leverage the inconsistent distortion characteristic of unbiased watermark $\delta$-reweight and address the shortcomings of traditional robust watermark methods that cannot detect modification effectively. Based on the inconsistent distortion, we introduce a modification detection method called IDD, which detects inconsistent tokens beside modified tokens. We propose an improved watermark detection method named drLLR, which enhances the robustness by dropping the score of inconsistent tokens. Experimental results demonstrate the effectiveness of IDD in detecting various types of modifications with high accuracy. At the same time, the drLLR method shows satisfactory performance in generated-text detection, which means we achieve effective dual detection capabilities by watermark. We hope this work can alleviate the potential threat of spoofing attacks and provide new references for designing LLM watermarks.

\bibliographystyle{unsrt}  
\bibliography{references}  

\begin{thebibliography}{10}

\bibitem{pan_risk_2023}
Yikang Pan, Liangming Pan, Wenhu Chen, Preslav Nakov, Min-Yen Kan, and willian Wang.
\newblock On the risk of misinformation pollution with large language models.
\newblock In Houda Bouamor, Juan Pino, and Kalika Bali, editors, {\em Findings of the Association for Computational Linguistics: {EMNLP} 2023}, pages 1389--1403. Association for Computational Linguistics, 2023.

\bibitem{kim2024llmsonlineemergingthreat}
Hanna Kim, Minkyoo Song, Seung~Ho Na, Seungwon Shin, and Kimin Lee.
\newblock When llms go online: The emerging threat of web-enabled llms.
\newblock {\em arXiv preprint arXiv:2410.14569}, 2024.

\bibitem{pmlr-v235-chakraborty24a}
Souradip Chakraborty, Amrit Bedi, Sicheng Zhu, Bang An, Dinesh Manocha, and Furong Huang.
\newblock Position: On the possibilities of {AI}-generated text detection.
\newblock In Ruslan Salakhutdinov, Zico Kolter, Katherine Heller, Adrian Weller, Nuria Oliver, Jonathan Scarlett, and Felix Berkenkamp, editors, {\em Proceedings of the 41st International Conference on Machine Learning}, volume 235 of {\em Proceedings of Machine Learning Research}, pages 6093--6115. PMLR, 21--27 Jul 2024.

\bibitem{Mitchelldetectgpt}
Eric Mitchell, Yoonho Lee, Alexander Khazatsky, Christopher~D. Manning, and Chelsea Finn.
\newblock Detectgpt: zero-shot machine-generated text detection using probability curvature.
\newblock In {\em Proceedings of the 40th International Conference on Machine Learning}, ICML'23. JMLR.org, 2023.

\bibitem{kirchenbauer2023reliability}
John Kirchenbauer, Jonas Geiping, Yuxin Wen, Manli Shu, Khalid Saifullah, Kezhi Kong, Kasun Fernando, Aniruddha Saha, Micah Goldblum, and Tom Goldstein.
\newblock On the reliability of watermarks for large language models.
\newblock {\em arXiv preprint arXiv:2306.04634}, 2023.

\bibitem{li2024statistical}
Xiang Li, Feng Ruan, Huiyuan Wang, Qi~Long, and Weijie~J Su.
\newblock A statistical framework of watermarks for large language models: Pivot, detection efficiency and optimal rules.
\newblock {\em arXiv preprint arXiv:2404.01245}, 2024.

\bibitem{Wu2023ASO}
Junchao Wu, Shu Yang, Runzhe Zhan, Yulin Yuan, Derek~F. Wong, and Lidia~S. Chao.
\newblock A survey on llm-generated text detection: Necessity, methods, and future directions.
\newblock {\em arXiv preprint arXiv:2310.14724}, 2023.

\bibitem{kamaruddin_review_2018}
Nurul~Shamimi Kamaruddin, Amirrudin Kamsin, Lip~Yee Por, and Hameedur Rahman.
\newblock A review of text watermarking: Theory, methods, and applications.
\newblock {\em {IEEE} Access}, 6:8011--8028, 2018.
\newblock Conference Name: {IEEE} Access.

\bibitem{yoo_advancing_2024}
{KiYoon} Yoo, Wonhyuk Ahn, and Nojun Kwak.
\newblock Advancing beyond identification: Multi-bit watermark for large language models.
\newblock In Kevin Duh, Helena Gomez, and Steven Bethard, editors, {\em Proceedings of the 2024 Conference of the North American Chapter of the Association for Computational Linguistics: Human Language Technologies (Volume 1: Long Papers)}, pages 4031--4055. Association for Computational Linguistics, 2024.

\bibitem{yang2022tracing}
Xi~Yang, Jie Zhang, Kejiang Chen, Weiming Zhang, Zehua Ma, Feng Wang, and Nenghai Yu.
\newblock Tracing text provenance via context-aware lexical substitution.
\newblock In {\em Proceedings of the AAAI Conference on Artificial Intelligence}, volume~36, pages 11613--11621, 2022.

\bibitem{kirchenbauer23a}
John Kirchenbauer, Jonas Geiping, Yuxin Wen, Jonathan Katz, Ian Miers, and Tom Goldstein.
\newblock A watermark for large language models.
\newblock In Andreas Krause, Emma Brunskill, Kyunghyun Cho, Barbara Engelhardt, Sivan Sabato, and Jonathan Scarlett, editors, {\em Proceedings of the 40th International Conference on Machine Learning}, volume 202 of {\em Proceedings of Machine Learning Research}, pages 17061--17084. PMLR, 23--29 Jul 2023.

\bibitem{zhao2024provable}
Xuandong Zhao, Prabhanjan~Vijendra Ananth, Lei Li, and Yu-Xiang Wang.
\newblock Provable robust watermarking for {AI}-generated text.
\newblock In {\em The Twelfth International Conference on Learning Representations}, 2024.

\bibitem{lee2024wrote}
Taehyun Lee, Seokhee Hong, Jaewoo Ahn, Ilgee Hong, Hwaran Lee, Sangdoo Yun, Jamin Shin, and Gunhee Kim.
\newblock Who wrote this code? watermarking for code generation.
\newblock In Lun-Wei Ku, Andre Martins, and Vivek Srikumar, editors, {\em Proceedings of the 62nd Annual Meeting of the Association for Computational Linguistics (Volume 1: Long Papers)}, pages 4890--4911, Bangkok, Thailand, August 2024. Association for Computational Linguistics.

\bibitem{wu2024dip}
Yihan Wu, Zhengmian Hu, Junfeng Guo, Hongyang Zhang, and Heng Huang.
\newblock A resilient and accessible distribution-preserving watermark for large language models.
\newblock In {\em Forty-first International Conference on Machine Learning}, 2024.

\bibitem{kuditipudi2024robust}
Rohith Kuditipudi, John Thickstun, Tatsunori Hashimoto, and Percy Liang.
\newblock Robust distortion-free watermarks for language models.
\newblock {\em Transactions on Machine Learning Research}, 2024.

\bibitem{pmlr-v247-christ24a}
Miranda Christ, Sam Gunn, and Or~Zamir.
\newblock Undetectable watermarks for language models.
\newblock In Shipra Agrawal and Aaron Roth, editors, {\em Proceedings of Thirty Seventh Conference on Learning Theory}, volume 247 of {\em Proceedings of Machine Learning Research}, pages 1125--1139. PMLR, 30 Jun--03 Jul 2024.

\bibitem{hu2024unbiased}
Zhengmian Hu, Lichang Chen, Xidong Wu, Yihan Wu, Hongyang Zhang, and Heng Huang.
\newblock Unbiased watermark for large language models.
\newblock In {\em The Twelfth International Conference on Learning Representations}, 2024.

\bibitem{pang2024freelunchllmwatermarking}
Pang Qi, Hu~Shengyuan, Zheng Wenting, and Smith Virginia.
\newblock No free lunch in llm watermarking: Trade-offs in watermarking design choices.
\newblock {\em arXiv preprint arXiv:2402.16187}, 2024.

\bibitem{gloaguen2024discoveringcluesspoofedlm}
Thibaud Gloaguen, Nikola Jovanović, Robin Staab, and Martin Vechev.
\newblock Discovering clues of spoofed lm watermarks.
\newblock {\em arXiv preprint arXiv:2410.02693}, 2024.

\bibitem{singh2023newevaluationmetricscapture}
Karanpartap Singh and James Zou.
\newblock New evaluation metrics capture quality degradation due to llm watermarking, 2023.

\bibitem{jin2019pubmedqa}
Qiao Jin, Bhuwan Dhingra, Zhengping Liu, William Cohen, and Xinghua Lu.
\newblock Pubmedqa: A dataset for biomedical research question answering.
\newblock In {\em Proceedings of the 2019 Conference on Empirical Methods in Natural Language Processing and the 9th International Joint Conference on Natural Language Processing (EMNLP-IJCNLP)}, pages 2567--2577, 2019.

\bibitem{zhang2022opt}
Susan Zhang, Stephen Roller, Naman Goyal, Mikel Artetxe, Moya Chen, Shuohui Chen, Christopher Dewan, Mona Diab, Xian Li, Xi~Victoria Lin, et~al.
\newblock Opt: Open pre-trained transformer language models.
\newblock {\em arXiv preprint arXiv:2205.01068}, 2022.

\end{thebibliography}






\end{document}